\documentclass[amsmath,amssymb]{revtex4}
\usepackage{graphicx}% Include figure files
\usepackage{amscd,amsmath,amsthm,amsfonts,amssymb}

\def\PT{$\cal{PT}$}
\def\[{\begin{equation}}
\def\]{\end{equation}}

\begin{document}
\title{Analytical construction of soliton families in one- and two-dimensional nonlinear Schr\"odinger equations with non-parity-time-symmetric complex potentials}
\author{Jianke Yang}
\affiliation{Department of Mathematics and Statistics, University of Vermont, Burlington, VT 05405, U.S.A}

\begin{abstract}
The existence of soliton families in non-parity-time-symmetric complex potentials remains poorly understood, especially in two spatial dimensions. In this article, we analytically investigate the bifurcation of soliton families from linear modes in one- and two-dimensional nonlinear Schr\"odinger equations with localized Wadati-type non-parity-time-symmetric complex potentials. By utilizing the conservation law of the underlying non-Hamiltonian wave system, we convert the complex soliton equation into a new real system. For this new real system, we perturbatively construct a continuous family of low-amplitude solitons bifurcating from a linear eigenmode to all orders of the small soliton amplitude. Hence, the emergence of soliton families in these non-parity-time-symmetric complex potentials is analytically explained. We also compare these analytically constructed soliton solutions with high-accuracy numerical solutions in both one and two dimensions, and the asymptotic accuracy of these perturbation solutions is confirmed.
\end{abstract}

\maketitle

\section{Introduction} \label{sec:intro}
Nonlinear wave phenomena in parity-time (\PT) symmetric systems have been under intensive studies in the past decade (see \cite{Konotop_review,Kivshar_review,PT_book} for reviews). Although the concept of \PT symmetry originated from non-Hermitian quantum mechanics \cite{Bender1998,Ali,Benderbook}, it was the interpretation of \PT symmetry as balanced gain and loss that made it flourish in optics and many other branches of physics \cite{Muss2008,Konotop_review,Kivshar_review,PT_book,Benderbook}. \PT symmetric systems are important for at least two reasons. From the intellectual point of view, these systems are the first reported non-Hamiltonian systems that, despite the gain and loss, exhibit many properties of Hamiltonian systems --- such as all-real linear spectra and continuous families of solitons \cite{Konotop_review,Kivshar_review,PT_book,Bender1998,Ali,Benderbook}. From the practical point of view, \PT symmetry has inspired many interesting applications, such as the coherent perfect absorber laser \cite{CPA1,CPA2,CPA3} and single-mode \PT lasers \cite{PTlaser1,PTlaser2}. While applications of \PT symmetry are still developing, its peculiar Hamiltonian-like phenomena, such as the existence of all-real spectrum and continuous families of solitons, have already been understood from a mathematical point of view
\cite{Ali,Benderbook,Konotop_review,Kivshar_review}. In particular, this understanding relies entirely on the \PT symmetry.

In the past few years, it was discovered that certain non-\PT-symmetric non-Hamiltonian systems also share properties of Hamiltonian systems. For example, the linear Schr\"odinger operator with certain types of non-\PT-symmetric complex potentials could still admit all-real spectra %\cite{Cannata98,Miri2013,Tsoy,NixonYang2016,Yang2018}.
\cite{Cannata98,Miri2013,Tsoy,NixonYang2016}. In addition, the one- and two-dimensional NLS equations with Wadati-type non-\PT-symmetric complex potentials could still admit continuous families of solitons \cite{Tsoy,Konotop2014,myPRA2D,NixonYangSAMP}. Furthermore, in the NLS equations with Wadati-type non-\PT-symmetric potentials, the linear-stability spectra of solitons still exhibit the quartet eigenvalue symmetry that is typical of Hamiltonian systems \cite{NixonYangPLA2016}. In a generic non-\PT-symmetric non-Hamiltonian system, none of these properties would hold. Thus, why these Hamiltonian-like properties arise in certain types of non-\PT-symmetric non-Hamiltonian systems is an intriguing theoretical question. While the all-real spectra of certain non-\PT-symmetric complex potentials can be explained by techniques such as supersymmetry and pseudo-Hermiticity \cite{Cannata98,Miri2013,Tsoy,NixonYang2016}, analytical explanations for the other properties associated with nonlinear non-\PT-symmetric systems remain elusive.

This article is concerned with the question of why the NLS equations with Wadati-type non-\PT-symmetric complex potentials could still admit continuous families of solitons. This phenomenon is peculiar, since these non-\PT-symmetric systems are non-Hamiltonian due to the presence of gain and loss, and solitons in non-Hamiltonian systems are generically isolated and do not exist as continuous families due to the double balancing requirement of nonlinearity with dispersion and gain with loss \cite{Akhmedievbook}. Numerical evidence to support this generic behavior in a non-\PT-symmetric system can be found in \cite{Panos_PTbook}, and a more mathematical reason for it can be found in \cite{YangPLA2014}. In view of this generic behavior and in the absence of \PT symmetry, why soliton families could appear in the NLS equations with special Wadati-type non-\PT-symmetric potentials is a deep mathematical mystery. It is physically meaningful for us to add that, unlike \PT-symmetric potentials where the spatial gain and loss distributions must be balanced in an exact anti-symmetric way, the Wadati potentials allow the gain and loss distributions to be arbitrary, which could potentially accommodate more realistic non-Hamiltonian physical systems in optics and beyond. A physical setup to realize Wadati potentials in a coherent atomic system has been proposed in \cite{Huang}.

In the one-dimensional (1D) case, some analytical understanding on this question has been provided in \cite{Konotop2014,NixonYangSAMP}. In \cite{Konotop2014}, Konotop and Zezyulin discovered a constant of motion for the underlying soliton equation with Wadati potentials. Combining this constant of motion with a shooting argument, the authors gave a plausible, but not definitive, explanation for these soliton families. In \cite{NixonYangSAMP}, the authors used this constant of motion to convert the original second-order complex soliton equation into a second-order real equation for the amplitude of the soliton. From this real soliton-amplitude equation, it was shown that continuous families of solitons bifurcating from linear modes could be constructed perturbatively. One drawback of this treatment in \cite{NixonYangSAMP} is that this real amplitude equation has some sign ambiguity in front of a square root term, which can cause technical complications. Another drawback, which is more serious, is that this treatment cannot be generalized to two and higher spatial dimensions.

In the 2D case, while soliton families in the 2D NLS equations with separable Wadati-type non-\PT-symmetric potentials were briefly mentioned on numerical grounds in \cite{myPRA2D}, there has been absolutely no analytical explanation for this phenomenon yet, except that a conservation law for the underlying non-Hamiltonian 2D equation was reported in that same article. Note that in this 2D case, the shooting argument of \cite{Konotop2014} no longer applies. In addition, the real-amplitude-equation treatment of \cite{NixonYangSAMP} also fails. Thus, new approaches need to be developed to analytically explain these 2D soliton families.

We would like to mention that continuous families of solitons in the 1D NLS equation perturbed by non-\PT-symmetric potentials more general than the Wadati-type were also reported by Kominis et al. \cite{Panos2019} through Melnikov's perturbation method. Since the authors' analysis was carried out only to the first order of the perturbation series, we suspect that those soliton families in non-Wadati potentials are valid only to the first order of the perturbation theory, but not to higher orders. If so, then those ``soliton families" would be just approximate solutions, but not true solitons. This suspicion makes it more imperative to analytically explain the existence of soliton families in non-\PT-symmetric Wadati potentials, since such analytical understanding could shed light on the nature of ``soliton families" reported in \cite{Panos2019} for non-Wadati potentials.

In this article, we analytically investigate the bifurcation of soliton families from linear modes in the 1D and 2D NLS equations with non-\PT-symmetric Wadati-type localized potentials through a new perturbative treatment. Utilizing the constant of motion of the underlying soliton equation, we convert this complex soliton equation into a new real system. The advantage of this new real system is that it allows us to analytically construct low-amplitude soliton families perturbatively to all orders of the amplitude in both one and two dimensions. Hence, soliton families in these 1D and 2D non-\PT-symmetric systems are analytically established. The reason this construction can be pursued to all orders is that the linear operator of these perturbation equations possesses two localized functions in its kernel, while the associated adjoint operator contains a single localized or bounded function in its kernel. These kernel structures, together with the phase invariance of solitons, ensure that at each order, the Fredholm condition for localized perturbation solutions can always be satisfied. Hence, we can construct a low-amplitude soliton solution, as a perturbation series to all orders, at each propagation constant in a continuous interval bordering the linear eigenmode of the potential. In other words, a soliton family bifurcating from a linear mode is derived in the underlying non-\PT-symmetric non-Hamiltonian system. These analytically constructed perturbation-series solutions for the soliton families are also compared to direct numerical solutions, and the asymptotic accuracy of these perturbation series solutions is confirmed.

\section{Construction of soliton families in the 1D case}
We first consider the 1D NLS equation
\[ \label{NLS1D}
iU_{t}+U_{xx}+V(x)U+\sigma |U|^2U=0
\]
with a non-\PT-symmetric Wadati potential
\[ \label{V}
V(x)=g^2(x)+{\rm{i}}g'(x),
\]
where $g(x)$ is an asymmetric real function that is differentiable everywhere, the prime represents differentiation, and $\sigma=\pm 1$ is the sign of cubic nonlinearity. Since $g(x)$ is real and asymmetric, $V^*(-x)\ne V(x)$, i.e., the complex potential $V(x)$ is non-\PT-symmetric \cite{Muss2008,Konotop_review,Kivshar_review,PT_book}. Potentials of this form appeared in Wadati's investigation of complex potentials with real spectra \cite{Wadati}, and are thus sometimes referred to as the Wadati potentials in the literature. In the optical context, the complex potential $V(x)$ in Eq. (\ref{NLS1D}) corresponds to the complex refractive index of the medium, where the imaginary part of $V(x)$, i.e., Im($V$),  describes the spatial gain and loss distributions, with regions of $\mbox{Im}(V)>0$ being lossy and regions of $\mbox{Im}(V)<0$ being gain \cite{Muss2008,Konotop_review,Kivshar_review,PT_book}. In this physical setting, since the function $g(x)$ in the Wadati potential (\ref{V}) can be arbitrary, this complex potential then can accommodate optical systems with arbitrary gain and loss distributions. The main constraint of the Wadati potential is that, the real refractive index profile of the medium, as described by the real part of the complex potential Re($V$), should be designed accordingly as $g^2(x)$. But this requirement on the real refractive index profile can be readily met given the sophisticated refractive-index engineering technology that is currently widely available.

An important property of the NLS equation (\ref{NLS1D}) with Wadati potentials is that, although this equation is non-Hamiltonian due to the complex potential, it admits a conservation law
\[ \label{QJ}
Q_t+J_x=0,
\]
where
\begin{equation}
Q=-U^*({\rm{i}}U_x-gU), \quad J=|U_x+{\rm{i}}gU|^2+{\rm{i}}U^*U_t+\frac{\sigma}{2} |U|^4,
\end{equation}
and the asterisk `*' represents complex conjugation. This conservation law is a special case of the more general conservation law reported in \cite{myPRA2D} for the 2D NLS equation with a separable Wadati-type potential.

Solitons in Eq. (\ref{NLS1D}) are of the form
\[ \label{Uu}
U(x,t)=e^{{\rm{i}}\mu t}u(x),
\]
where $\mu$ is a real propagation constant, and $u(x)$ is a localized function satisfying the soliton equation
\[ \label{ueq}
u_{xx}+(g^2+{\rm{i}}g')u-\mu u +\sigma |u|^2u=0.
\]
Notice that this complex soliton equation is phase-invariant, i.e., if $u(x)$ is a solution, so is $e^{{\rm{i}}\theta}u(x)$, where $\theta$ is an arbitrary real constant. Substituting the soliton solution (\ref{Uu}) into the conservation law (\ref{QJ}), we get $dJ/dx=0$, where
\[\label{eqJ}
J(x)=|u_x+{\rm{i}}gu|^2-\mu |u|^2 +\frac{\sigma}{2} |u|^4.
\]
Since solitons decay to zero as $x\to \pm\infty$, we see that $J(x)=0$, which is a constant of motion for the soliton equation (\ref{ueq}). This constant of motion is equivalent to the one reported in Ref.~\cite{Konotop2014} for the same equation (\ref{ueq}).

Soliton families in Eq. (\ref{ueq}), parameterized by the propagation constant $\mu$, for non-\PT-symmetric Wadati potentials were reported numerically in \cite{Tsoy}, and studied analytically in \cite{Konotop2014,NixonYangSAMP,YangPLA2014} with limited success. In particular, the perturbative construction of small-amplitude soliton families as proposed in \cite{YangPLA2014,NixonYangSAMP} exhibits some difficulties. The perturbative construction in \cite{YangPLA2014} was based on the complex soliton equation (\ref{ueq}). The difficulty with this construction, as explained in \cite{YangPLA2014}, is that each order of the perturbation series creates a nontrivial condition which needs to be satisfied, and it is almost impossible to prove that all those infinite number of conditions would hold. The perturbative construction in \cite{NixonYangSAMP} was based on a real second-order equation for the amplitude $|u(x)|$ of the soliton, and this real amplitude equation was derived from the original complex equation (\ref{ueq}) with the help of the above constant of motion $J(x)=0$.
This latter construction removed those infinite number of nontrivial conditions of the former, and thus made the perturbative construction possible, at least in principle. But it does create some technical difficulties.
For example, this reduced amplitude equation contains a square root term, whose sign can be ambiguous and cause technical complications. To remove this ambiguity, some technical assumptions had to be imposed in \cite{NixonYangSAMP}. A more serious problem with this latter treatment is that it does not work for the 2D case. In other words, in two (and higher) spatial dimensions, we will not be able to convert the original complex soliton equation into a single real equation for the amplitude of the soliton.

In this section, we will develop a new perturbative construction of low-amplitude soliton families in Eq. (\ref{ueq}), which can be easily pursued to all orders of the perturbation series. More importantly, this new 1D treatment can be readily generalized to the 2D case.

For the technical convenience of our perturbative construction, we will assume that the Wadati potential (\ref{V}) is localized in space, i.e., the real function $g(x)$ in this potential will be assumed to be localized. This assumption of locality on the potential has two main benefits. One is that such a Wadati potential often admits a discrete real eigenvalue \cite{Tsoy,NixonYang2016}, which is the starting point of our perturbative calculation. The other is that under this locality assumption, the eigenfunction associated with this discrete real eigenvalue of the potential features simple and explicit exponential decay at large distances. These explicit decay rates of the eigenfunction facilitate our derivation and understanding of the kernels for the linearization operator and its adjoint in the upcoming section \ref{sec_Kernel}. If the Wadati potential (\ref{V}) is not localized (for instance, unbounded) but still admits a discrete real eigenvalue, then the analysis of this section can still go through, because the kernel structures of the linearization operator and its adjoint to be established in Sec.~\ref{sec_Kernel} would still remain valid. However, if the Wadati potential (\ref{V}) is periodic, then the situation would be different. In this case, the periodic potential does not admit any discrete real eigenvalues. Instead, the spectrum of the potential comprises Bloch bands. Low-amplitude solitons, if any, would have to bifurcate out from edges of these Bloch bands as envelope solitons \cite{Pelinovskybook}. The analytical calculation of soliton bifurcation from Bloch-band edges in a periodic Wadati potential would be very different from the one to be developed in this section, and it will be left for future studies.

\subsection{A new real system for solitons and its perturbation expansion}

It can be checked that the original complex soliton equation (\ref{ueq}) is equivalent to two real equations --- one is that the real part of (\ref{ueq}) is zero, and the other is $J=0$, where $J$ is given in Eq. (\ref{eqJ}).
The first real equation comes directly from (\ref{ueq}), and the second one is the constant of motion discussed below Eq. (\ref{eqJ}). To see these two real equations combined could also reproduce the original complex equation (\ref{ueq}), we only need to notice that $dJ/dx$ is equal to the real part of the product between $u_x^*-igu^*$ and the left side of the complex soliton equation (\ref{ueq}). Thus, if $J=0$ and the real part of (\ref{ueq}) is zero, then the imaginary part of (\ref{ueq}) needs to be zero as well.

Expressing $u(x)$ as
\[ \label{upq}
u(x)=p(x)+{\rm{i}}q(x),
\]
where $p(x)$ and $q(x)$ are the real and imaginary parts of the complex function $u(x)$, these two real equations for solitons are
\begin{eqnarray}
&& p_{xx}+(g^2-\mu)p-g'q+\sigma(p^2+q^2)p=0,    \label{peq}\\
&&(p_x-gq)^2+(q_x+gp)^2-\mu(p^2+q^2)+\frac{\sigma}{2}(p^2+q^2)^2=0.  \label{qeq}
\end{eqnarray}
This system of two real equations will be the one we use to analytically calculate soliton families. It is important to notice that this is a third-order real system, which contrasts the original soliton equation (\ref{ueq}), which is a fourth-order real system when that complex equation is split into two real second-order equations for $p(x)$ and $q(x)$. This third-order real system also contrasts the second-order real system we derived in Ref.~\cite{NixonYangSAMP} for the amplitude $|u(x)|$ of the soliton.

Now, we perturbatively construct a continuous family of low-amplitude solitons bifurcating from a linear discrete eigenmode of a localized Wadati potential. Suppose the Schr\"odinger operator $\partial_{xx}+V(x)$ with a localized Wadati potential (\ref{V}) admits a discrete real eigenvalue $\mu_0$, whose eigenfunction is $\phi(x)+{\rm{i}}\psi(x)$, where $\phi(x)$ and $\psi(x)$ are localized real functions. Then,
\[ \label{eigcomp}
\left(\partial_{xx}+g^2+{\rm{i}}g'\right) (\phi+{\rm{i}}\psi)=\mu_0 (\phi+{\rm{i}}\psi).
\]
The existence of such a real eigenvalue is common in a Wadati potential. For instance, it was shown in \cite{Tsoy} that if $g(x)$ is a single-humped localized real function, then the spectrum of the corresponding Wadati potential is strictly real. In the more general case, it was shown in \cite{NixonYang2016} that eigenvalues in a Wadati potential always come as complex-conjugate pairs and are thus often real.  %\cite{Tsoy,NixonYang2016,Yang2018}.
Because this potential is assumed to be localized, its discrete real eigenvalue $\mu_0$ must be positive, i.e., $\mu_0>0$.

Bifurcating from this linear discrete eigenmode, we seek a low-amplitude soliton at each real propagation constant $\mu$ near $\mu_0$, and this soliton can be expanded into the following perturbation series,
\begin{eqnarray}
&& p(x; \mu) = \epsilon^{1/2}\left[ p_0(x)+\epsilon p_1(x) +\epsilon^2 p_2(x) +\cdots\right],   \label{pexpand}\\
&& q(x; \mu) = \epsilon^{1/2}\left[ q_0(x)+\epsilon q_1(x) +\epsilon^2 q_2(x) +\cdots\right],   \label{qexpand}
\end{eqnarray}
where $\epsilon=\mu-\mu_0$ and is assumed to be small positive (so that $\epsilon^{1/2}$ is real). This means that we assume the bifurcation is to the right side of $\mu_0$, i.e., $\mu>\mu_0$. As we will see in later text [see Eq. (\ref{c02})], this rightward bifurcation can be induced by a proper choice on the sign of nonlinearity $\sigma$. If this sign of nonlinearity is opposite of that choice, the soliton bifurcation will be to the left side of $\mu_0$. In that case, we can define $\epsilon=\mu_0-\mu$, and the rest of the perturbative calculation would be very similar.

Substituting the above perturbation expansion into the real system (\ref{peq})-(\ref{qeq}), we get a sequence of real equations for the functions $(p_k, q_k)$. The equations for $(p_0, q_0)$ are
\begin{eqnarray}
&& (\partial_{xx}+g^2-\mu_0)p_0-g'q_0=0,  \label{p0q01} \\
&& (p_{0x}-gq_0)^2+(q_{0x}+gp_0)^2-\mu_0(p_0^2+q_0^2)=0.  \label{p0q02}
\end{eqnarray}
Even though this is a nonlinear system, it is scaling invariant, i.e., if $(p_0, q_0)$ is a solution, so is $(\alpha p_0, \alpha q_0)$, where $\alpha$ is an arbitrary real constant. Thus, this system is actually an eigenvalue problem in disguise and is equivalent to the linear complex eigenvalue problem (\ref{eigcomp}).
Its solution then is
\[ \label{p0q0}
\left[\begin{array}{c} p_0 \\ q_0 \end{array}\right]=c_0 \left[\begin{array}{c} \phi \\ \psi \end{array}\right],
\]
where $c_0$ is a real constant to be determined. Indeed, since $\phi+{\rm{i}}\psi$ is a solution to the linear eigenvalue problem (\ref{eigcomp}), the above $(p_0, q_0)$ then satisfy the original equations (\ref{peq})-(\ref{qeq}) to leading order, which are Eqs. (\ref{p0q01})-(\ref{p0q02}).

Utilizing the above $(p_0, q_0)$ solution, we find that the functions $(p_k, q_k)$ for $k\ge 1$ are governed by the following linear nonhomogeneous system of equations
\[ \label{Lpkqk}
{\cal L}\left[\begin{array}{c} p_k \\ q_k \end{array}\right]= \left[\begin{array}{c} f_k \\
g_k \end{array}\right],
\]
where
\[ \label{calL}
{\cal L}=\left[\begin{array}{ll} \partial_{xx}+g^2-\mu_0  & -g' \\
(\phi'-g\psi)\partial_x+g(\psi'+g\phi)-\mu_0\phi & (\psi'+g\phi)\partial_x-g(\phi'-g\psi)-\mu_0\psi \end{array}\right],
\]
\[ \label{f1g1}
\left[\begin{array}{c} f_1 \\ g_1 \end{array}\right]= c_0 \left[\begin{array}{l} \phi -\sigma c_0^2 (\phi^2+\psi^2)\phi \\
\frac{1}{2}(\phi^2+\psi^2)-\frac{1}{4}\sigma c_0^2 (\phi^2+\psi^2)^2 \end{array}\right],
\]
\[ \label{f2g2}
\left[\begin{array}{c} f_2 \\ g_2 \end{array}\right]= \left[\begin{array}{l} \left(1 -3\sigma p_0^2 -\sigma q_0^2\right)p_1-2\sigma p_0q_0q_1 \\ \frac{1}{2c_0}\left[2\left(1-\sigma p_0^2-\sigma q_0^2\right)(p_0p_1+q_0q_1)+\mu_0(p_1^2+q_1^2)-(p_{1x}-gq_1)^2-(q_{1x}+gp_1)^2)\right] \end{array}\right],
\]
\[ \label{fkgk}
\left[\begin{array}{c} f_k \\ g_k \end{array}\right]= \left[\begin{array}{cc} {\cal M}_{11} & {\cal M}_{12} \\
{\cal M}_{21} & {\cal M}_{22} \end{array}\right]\left[\begin{array}{c}
p_{k-1} \\ q_{k-1}\end{array}\right] +\left[\begin{array}{c} {\cal N}_{k}^{[1]} \\ {\cal N}_{k}^{[2]}\end{array}\right],    \quad k\ge 3,
\]
the matrix elements ${\cal M}_{ij}$ are $k$-independent and given by the formulae
\begin{eqnarray*}
&& {\cal M}_{11}=1-3\sigma p_0^2-\sigma q_0^2, \\
&& {\cal M}_{12}=-2\sigma p_0q_0, \\
&& {\cal M}_{21}=\frac{1}{c_0}\left[p_0\left( 1-\sigma p_0^2-\sigma q_0^2\right)+\mu_0p_1-(p_{1x}-gq_1)\partial_x-g(q_{1x}+gp_1)\right],\\
&& {\cal M}_{22}=\frac{1}{c_0}\left[q_0\left( 1-\sigma p_0^2-\sigma q_0^2\right)+\mu_0q_1+g(p_{1x}-gq_1)-(q_{1x}+gp_1)\partial_x\right],
\end{eqnarray*}
and ${\cal N}_{k}^{[1]}, {\cal N}_{k}^{[2]}$ are functions which depend only on $k$, $p_0, p_1, \dots, p_{k-2}$, $q_0, q_1, \dots, q_{k-2}$ and $g(x)$. For example, when $k=3$,
\begin{eqnarray*}
&& {\cal N}_{3}^{[1]}=-\sigma\left(3p_0p_1^2+2p_1q_0q_1+p_0q_1^2\right), \\
&& {\cal N}_{3}^{[2]}=\frac{1}{2c_0}\left[(p_1^2+q_1^2)\left(1-\sigma p_0^2-\sigma q_0^2\right)-2\sigma (p_0p_1+q_0q_1)^2\right].
\end{eqnarray*}

Next, we will show that we can solve the linear nonhomogeneous equations (\ref{Lpkqk}) and obtain localized solutions $(p_k, q_k)$ for all $k$, using the Fredholm alternative method.

\subsection{Kernel structures of the linear operator and its adjoint operator} \label{sec_Kernel}
The key to solving linear nonhomogeneous equations (\ref{Lpkqk}) by the Fredholm alternative method is to understand the kernel structures of the linear operator ${\cal L}$ and its adjoint operator ${\cal L}^A$. Under the inner product of
\[
\langle F, G \rangle \equiv \int_{-\infty}^\infty [F(x)]^T  \hspace{0.01cm} G(x) \hspace{0.06cm} {\rm{d}} x,
\]
where the superscript `$T$' represents the transpose of a vector or matrix, the adjoint operator of ${\cal L}$ is
\[ \label{calLA}
{\cal L}^A=\left[\begin{array}{ll} \partial_{xx}+g^2-\mu_0  & -\partial_x(\phi'-g\psi)+g(\psi'+g\phi)-\mu_0\phi \\
-g'  & -\partial_x(\psi'+g\phi)-g(\phi'-g\psi)-\mu_0\psi \end{array}\right].
\]

First, we consider the kernel structure of operator ${\cal L}$. It is easy to check that this kernel contains the following two localized functions
\[ \label{K1K2}
K_1\equiv \left[\begin{array}{c} \phi \\ \psi \end{array}\right], \quad
K_2\equiv \left[\begin{array}{c} -\psi \\ \phi \end{array}\right],
\]
where
\[
{\cal L}K_1={\cal L}K_2=0.
\]
Indeed, ${\cal L}K_1=0$ is equivalent to the complex linear eigenvalue equation (\ref{eigcomp}), and ${\cal L}K_2=0$ is equivalent to this complex eigenvalue equation with the eigenfunction changing from $\phi+{\rm{i}}\psi$ to ${\rm{i}}(\phi+{\rm{i}}\psi)$, which clearly remains an eigenfunction. Another way to understand these kernel functions is that, the first kernel function $K_1$ is induced by the scaling invariance of the complex linear eigenvalue equation (\ref{eigcomp}), and the second kernel function $K_2$ is induced by the phase invariance of that same equation.

It is clear that ${\cal L}$ is a third-order differential operator. Thus, the system ${\cal L}K=0$ admits one more linearly independent solution $K_3$ in addition to $K_1$ and $K_2$. This third solution is obviously unbounded in space. Indeed, since $\phi+{\rm{i}}\psi$ is the eigenfunction of the Schr\"odinger operator with a localized potential at the positive eigenvalue $\mu_0$ [see Eq. (\ref{eigcomp})], both $\phi(x)$ and $\psi(x)$ decay exponentially at the rate of $e^{-\sqrt{\mu_0}\hspace{0.03cm}|x|}$ when $x\to \pm \infty$. Then, converting the system ${\cal L}K=0$ into a system of three first-order equations and using Abel's formula, we can show that this third solution $K_3(x)$ grows exponentially at the rate of $e^{\sqrt{\mu_0}\hspace{0.03cm}|x|}$ when $x\to \pm \infty$.

Next, we consider the kernel structure of ${\cal L}^A$. Functions in this kernel can be derived from the functions in the kernel of ${\cal L}$. One way to do so is to first rewrite the equation ${\cal L}K=0$ with $K\equiv [K^{[1]},K^{[2]}]^T$ as a first-order system
\[ \label{YP}
\partial_xY=P(x)Y
\]
for $Y=[K^{[1]}, K^{[1]}_x, K^{[2]}]^T$, where $P(x)$ is a $3\times 3$ real matrix function. The fundamental matrix ${\cal Y}(x)$ of this first-order homogeneous system is given through the three solutions $K_1, K_2$ and $K_3$ of the original system ${\cal L}K=0$ as
\[
{\cal Y}=\left[\begin{array}{ccc} K_1^{[1]} & K_2^{[1]} & K_3^{[1]} \\
K_{1,x}^{[1]} & K_{2,x}^{[1]} & K_{3,x}^{[1]} \\
K_1^{[2]} & K_2^{[2]} & K_3^{[2]} \end{array}\right].
\]
The adjoint of the first-order system (\ref{YP}) is
\[ \label{YPA}
-\partial_x Y^A=P^TY^A,
\]
whose fundamental matrix is ${\cal Y}^A=({\cal Y}^{-1})^T$. Using the large-$x$ asymptotics of the $(K_1, K_2, K_3)$ solutions described in the previous paragraph, together with their Wronskian expression from Abel's formula, we can readily show that the third column of ${\cal Y}^A$ is localized with its second component decaying at the rate of $e^{-\sqrt{\mu_0}\hspace{0.03cm}|x|}$ at large $|x|$, while the first and second columns of ${\cal Y}^A$ are unbounded with their second components growing at the rate of $e^{\sqrt{\mu_0}\hspace{0.03cm}|x|}$ at large $|x|$.

The adjoint first-order system (\ref{YPA}) has a simple connection with the original adjoint system ${\cal L}^AK^A=0$. Specifically, if $Y^A=[Y^{A[1]}, Y^{A[2]}, Y^{A[3]}]^T$, then $K^{A}=[Y^{A[2]}, Y^{A[3]}/(\psi'+g\phi)]^T$. Using this connection, we see that the kernel of ${\cal L}^A$ contains a single localized function, which we denote as
\[ \label{K0A}
K_0^A=\left[\begin{array}{c} \phi^A \\ \psi^A \end{array}\right],
\]
where ${\cal L}^AK_0^A=0$. This $K_0^A$ is obtained from the third column of ${\cal Y}^A$; so $\phi^A(x)$ decays at the rate of $e^{-\sqrt{\mu_0}\hspace{0.03cm}|x|}$ when $x\to \pm \infty$. Regarding the decay rate of $\psi^A(x)$, using dominant balance on the second equation of the adjoint system ${\cal L}^AK^A=0$, we can show that $\psi^A(x)$ decays at the same rate of $g(x)$ for large $|x|$. The other two functions in the kernel of ${\cal L}^A$ are obtained from the first and second columns of ${\cal Y}^A$ and are thus both unbounded. More specifically, their first components grow at the rate of $e^{\sqrt{\mu_0}\hspace{0.03cm}|x|}$, and their second components grow at the rate of $e^{2\hspace{-0.03cm}\sqrt{\mu_0}\hspace{0.03cm}|x|}$, when $x\to \pm \infty$.

\subsection{The Fredholm solvability condition} \label{sec:Fred}
Utilizing the above kernel structures of operators ${\cal L}$ and ${\cal L}^A$, we can solve the linear nonhomogeneous equations (\ref{Lpkqk}) and obtain a localized solution $(p_k, q_k)$ for all $k$. To do so, we will use the Fredholm solvability condition, which will be explained in this subsection.

First, we notice that $f_k$ on the right side of the nonhomogeneous equations (\ref{Lpkqk}) is localized, and its decay rate at large $|x|$ is $e^{-\sqrt{\mu_0}\hspace{0.03cm}|x|}$, multiplied by a certain polynomial function of $x$. In addition, $g_k$ on the right side of these equations is also localized, and its decay rate at large $|x|$ is $e^{-2\hspace{-0.03cm}\sqrt{\mu_0}\hspace{0.03cm}|x|}$, multiplied by another polynomial function of $x$. The reason for these decay rates of $(f_k, g_k)$ is that $p_n$ and $q_n$ in the expressions of $f_k$ and $g_k$ decay at the rate of $e^{-\sqrt{\mu_0}\hspace{0.03cm}|x|}$, multiplied by a polynomial function of $x$. These decay rates of $(p_n, q_n)$ can be seen from the $\epsilon$ expansions (\ref{pexpand})-(\ref{qexpand}) of solitons $(p, q)$, which decay at the rate of $e^{-\sqrt{\mu_0+\epsilon}\hspace{0.03cm}|x|}$ at large $|x|$. These decay rates of $(p_n, q_n)$ can also be seen from the equations (\ref{Lpkqk}) which determine them.

In view of the decay rates of $(f_k, g_k)$ on the right side of the linear nonhomogeneous equations (\ref{Lpkqk}), as well as the kernel structures of linear operators ${\cal L}$ and ${\cal L}^A$ delineated in the previous subsection, the Fredholm alternative theorem says that these nonhomogeneous equations (\ref{Lpkqk}) would admit a localized solution $(p_k, q_k)$ if and only if the nonhomogeneous term $(f_k, g_k)^T$ is orthogonal to the localized function $K_0^A$ in the kernel of ${\cal L}^A$, i.e.,
\[ \label{Fredcond}
\left\langle \left[\begin{array}{c}  \phi^A \\ \psi^A \end{array}\right], \hspace{0.05cm}
\left[\begin{array}{c}  f_k \\ g_k \end{array}\right]\right\rangle
=\int_{-\infty}^\infty \left(\phi^A f_k + \psi^A g_k\right) \hspace{0.04cm} {\rm{d}} x=0.
\]
The Fredholm alternative theorem was originally developed for compact operators (\cite{Brezis_book}, page 160), which is restrictive. But this theorem can be generalized to operators with closed range (\cite{Brezis_book}, page 46). In this article, we will not attempt to prove that our operator ${\cal L}$ has closed range. Instead, we will provide an elementary proof of this Fredholm alternative result below.

The necessity of the above condition (\ref{Fredcond}) for Eq. (\ref{Lpkqk}) to admit a localized solution can be derived quickly by taking the inner product of this equation with the localized function $K_0^A$ in the kernel of ${\cal L}^A$. To prove the sufficiency of this condition, we can first rewrite Eq. (\ref{Lpkqk}) as a first-order system
\[ \label{YF}
\partial_xY-P(x)Y=F,
\]
where $Y=[p_k, p_{k,x}, q_k]^T$, $F=[0, f_k, g_k/(\psi'+g\phi)]^T$, and $P(x)$ is the $3\times 3$ real matrix function in Eq. (\ref{YP}). The fundamental matrix ${\cal Y}$ for the first-order homogeneous system of (\ref{YF}) has been discussed before. Using this fundamental matrix and variation of parameters, we can derive the general solution to the nonhomogeneous system (\ref{YF}) as
\[
Y(x)={\cal Y}(x)\left( \mathbf{c} + \int_0^x \left[{\cal Y}^{A}(z)\right]^T\hspace{-0.06cm} F(z) \hspace{0.04cm} {\rm{d}} z\right),
\]
where $\mathbf{c}$ is a constant vector, and ${\cal Y}^{A}=({\cal Y}^{-1})^T$ is the fundamental matrix of the first-order adjoint system (\ref{YPA}). In view of this explicit solution formula for Eq. (\ref{YF}), as well as the large-$x$ asymptotics of fundamental matrices ${\cal Y}$ and ${\cal Y}^{A}$ described earlier, we can readily see that a localized solution $Y(x)$ can be obtained, through a proper choice of the third element of the $\mathbf{c}$ constant, if the following condition is met,
\[ \label{cond0}
\int_{-\infty}^\infty \left[{\cal Y}_3^{A}(x)\right]^T\hspace{-0.06cm} F(x) \hspace{0.04cm} {\rm{d}} x=0,
\]
where ${\cal Y}_3^{A}$ is the third column of ${\cal Y}^{A}$. This third column is connected to the localized function $K_0^A$ through a relation explained in the last paragraph of the previous subsection. Then, using the expression of $F$ given above, the above condition
(\ref{cond0}) reduces exactly to the Fredholm solvability condition (\ref{Fredcond}). Thus, the sufficiency of this Fredholm condition to guarantee the existence of a localized solution in Eq. (\ref{Lpkqk}) is directly proved.

\subsection{Construction of perturbation series to all orders} \label{sec:pert}
Now, we use the Fredholm solvability condition (\ref{Fredcond}) to determine a soliton solution $u(x; \mu)$ through the perturbation series (\ref{pexpand})-(\ref{qexpand}), to all orders of $\epsilon\equiv \mu-\mu_0$, at each $\mu$ value near $\mu_0$. These solutions then constitute a continuous family of solitons, parameterized by the propagation constant $\mu$, in the non-\PT-symmetric Wadati potential (\ref{V}).

We first consider Eq. (\ref{Lpkqk}) for $(p_1, q_1)$. Substituting the $(f_1, g_1)$ expressions (\ref{f1g1}) into the Fredholm solvability condition (\ref{Fredcond}) and simplifying, we see that Eq. (\ref{Lpkqk}) admits a localized solution $(p_1, q_1)$ if and only if the constant $c_0$ is selected as
\[ \label{c02}
c_0 =\pm \sqrt{ \frac{\int_{-\infty}^\infty \left[\phi \phi^A+\frac{1}{2}(\phi^2+\psi^2)\psi^A\right] \hspace{0.04cm} {\rm{d}}x}{\sigma \int_{-\infty}^\infty \left[ (\phi^2+\psi^2)\phi\phi^A + \frac{1}{4} (\phi^2+\psi^2)^2\psi^A\right] \hspace{0.04cm} {\rm{d}}x}}.
\]
In order for the quantity under the square root above to be positive, $\sigma$ must have the same sign as the ratio of the two integrals in the above formula. In other words, in order for the soliton bifurcation to appear for $\mu > \mu_0$, the nonlinearity must be of a certain sign. In this case, $c_0$ has two value choices which differ by a sign. But it is easy to see that these two sign choices in $c$ would simply lead to two soliton solutions $u(x)=p(x)+{\rm{i}}q(x)$ which also differ only by a sign. Since the $u(x)$ equation (\ref{ueq}) is phase-invariant, solutions differing by a sign are equivalent. Thus, we will just take the plus sign for $c_0$ below.

When $c_0$ is selected from the above formula (\ref{c02}), Eq. (\ref{Lpkqk}) admits a localized solution for $(p_1, q_1)$, which we denote as $(p_{1s}, q_{1s})$. However, since the kernel of the homogeneous operator ${\cal L}$ in Eq. (\ref{Lpkqk}) contains two localized functions $K_1$ and $K_2$ given in Eq. (\ref{K1K2}), the general localized solution $(p_1, q_1)$ to the linear nonhomogeneous equations (\ref{Lpkqk}) is then
\[ \label{p1q1}
\left[\begin{array}{c}  p_1 \\ q_1 \end{array}\right]=\left[\begin{array}{c}  p_{1s} \\ q_{1s} \end{array}\right]+ c_1 \left[\begin{array}{c} \phi \\ \psi \end{array}\right] +d_1 \left[\begin{array}{c} -\psi \\ \phi \end{array}\right],
\]
where $c_1$ and $d_1$ are two real constants.

It is important to recognize that the $d_1$ term above can be removed by phase invariance of the complex soliton solution $u(x)$. To see this more clearly, we put the above perturbation solutions together and get
\begin{eqnarray*}
&& u(x)=\epsilon^{1/2}\left[ c_0\phi +\epsilon (p_{1s}+c_1\phi-d_1\psi) +\rm{i}\left[c_0\psi + \epsilon (q_{1s}+c_1\psi+d_1\phi)\right]+O(\epsilon^2)\right] \\
&& \hspace{0.67cm} =\epsilon^{1/2}\left[ (c_0+\epsilon c_1+\rm{i}\epsilon d_1)(\phi+\rm{i}\psi)+\epsilon (p_{1s}+\rm{i}q_{1s})+O(\epsilon^2)\right]\\
&& \hspace{0.67cm} =\epsilon^{1/2}e^{\rm{i} \epsilon d_1/c_0}\left[ (c_0+\epsilon c_1)(\phi+\rm{i}\psi)+\epsilon (p_{1s}+\rm{i}q_{1s})+O(\epsilon^2)\right].
\end{eqnarray*}
Notice that the $d_1$ term only contributes a constant phase of order $\epsilon$ to the soliton solution $u(x)$. But $u(x)$ is phase-invariant. Thus, that $d_1$ term in (\ref{p1q1}) can be dropped and we can set
\[ \label{p1q12}
\left[\begin{array}{c}  p_1 \\ q_1 \end{array}\right]=\left[\begin{array}{c}  p_{1s} \\ q_{1s} \end{array}\right]+ c_1 \left[\begin{array}{c} \phi \\ \psi \end{array}\right]
\]
without loss of generality.

The $(p_1, q_1)$ solution in the above equation contains an unknown real constant $c_1$. This $c_1$ constant will be determined from the Fredholm solvability condition on the $(p_2, q_2)$ equations. The equations for $(p_2, q_2)$ are (\ref{Lpkqk}), where the nonhomogeneous terms $(f_2, g_2)$ are given in Eq. (\ref{f2g2}). Substituting the $(p_0, q_0)$ solutions (\ref{p0q0})
and $(p_1, q_1)$ solutions (\ref{p1q12}) into the $(f_2, g_2)$ expressions (\ref{f2g2}) and recalling that $(\phi, \psi)$ satisfy the equation (\ref{p0q02}), we find that the $(f_2, g_2)$ expressions (\ref{f2g2}) reduce to
\[ \label{f2g2b}
\left[\begin{array}{c} f_2 \\ g_2 \end{array}\right]=\left[\begin{array}{c} f_{2a} \\ g_{2a} \end{array}\right]+c_1 \left[\begin{array}{c} f_{2b} \\ g_{2b} \end{array}\right],
\]
where
\begin{eqnarray*}
&& f_{2a}=(1-3\sigma p_0^2-\sigma q_0^2)p_{1s}-2\sigma p_0q_0q_{1s}, \\
&& f_{2b}=(1-3\sigma p_0^2-\sigma q_0^2)\phi-2\sigma p_0q_0\psi, \\
&& g_{2a}=\frac{1}{2c_0}
\left[2(1-\sigma p_0^2-\sigma q_0^2)(p_0 p_{1s}+q_0q_{1s})+\mu_0(p_{1s}^2+q_{1s}^2)-(p_{1s, x}-gq_{1s})^2-(q_{1s,x}+gp_{1s})^2\right],\\
&& g_{2b}=\frac{1}{c_0}
\left[(1-\sigma p_0^2-\sigma q_0^2)(p_0 \phi+q_0 \psi)+\mu_0(p_{1s}\phi+q_{1s}\psi)  \right.  \\
&& \hspace{1.35cm} \left. -(p_{1s, x}-gq_{1s})(\phi_x-g\psi)-(q_{1s,x}+gp_{1s})(\psi_x+g\phi)\right],
\end{eqnarray*}
which are independent of the unknown constant $c_1$. Then, the Fredholm solvability condition (\ref{Fredcond}) at $k=2$ gives the formula for the constant $c_1$ as
\[ \label{c1}
c_1=-\frac{\int_{-\infty}^\infty \left(\phi^A f_{2a}+\psi^Ag_{2a}\right)\hspace{0.03cm} \hspace{0.04cm} {\rm{d}}x}{\int_{-\infty}^\infty \left(\phi^A f_{2b}+\psi^Ag_{2b}\right)\hspace{0.03cm} \hspace{0.04cm} {\rm{d}}x}.
\]

The rest of the perturbation calculations can then proceed to all orders as follows. When $c_{k-1}$ ($k\ge 2$) has been obtained, the $(p_{k-1}, q_{k-1})$ solutions are completely determined. Meanwhile, the solvability condition (\ref{Fredcond}) for $(p_k, q_k)$ is also satisfied, and thus there exists a localized solution which we denote as $(p_{k,s}, q_{k,s})$. The general localized solutions for $(p_k, q_k)$
can be written as
\[ \label{pkqk}
\left[\begin{array}{c}  p_k \\ q_k \end{array}\right]=\left[\begin{array}{c}  p_{k,s} \\ q_{k,s} \end{array}\right]+ c_k \left[\begin{array}{c} \phi \\ \psi \end{array}\right].
\]
The constant $c_k$ will be determined from the solvability condition for the $(p_{k+1}, q_{k+1})$ equations (\ref{Lpkqk}). Specifically, when the above $(p_k, q_k)$ solutions are inserted into the $(f_{k+1}, g_{k+1})$ formulae (\ref{fkgk}), it is easy to see that the solvability condition (\ref{Fredcond}) at $k+1$ is a linear equation for $c_k$, which we can easily solve to obtain the value of $c_k$ as
\begin{equation*}
c_k=-\frac{
\left\langle \left[\begin{array}{c}  \phi^A \\ \psi^A \end{array}\right], \hspace{0.05cm}
\left[\begin{array}{cc} {\cal M}_{11} & {\cal M}_{12} \\
{\cal M}_{21} & {\cal M}_{22} \end{array}\right]\left[\begin{array}{c}
p_{k,s} \\ q_{k,s}\end{array}\right]+\left[\begin{array}{c} {\cal N}_{k+1}^{[1]} \\ {\cal N}_{k+1}^{[2]}\end{array}\right]\right\rangle}
{ \left\langle \left[\begin{array}{c}  \phi^A \\ \psi^A \end{array}\right], \hspace{0.05cm}
\left[\begin{array}{cc} {\cal M}_{11} & {\cal M}_{12} \\
{\cal M}_{21} & {\cal M}_{22} \end{array}\right]\left[\begin{array}{c}
\phi \\ \psi \end{array}\right]\right\rangle}, \quad k\ge 2.
\end{equation*}
Utilizing the $(p_1, q_1)$ formula (\ref{p1q12}) and the fact that $(\phi, \psi)$ satisfy Eq.~(\ref{p0q02}), we can verify that the denominator in this $c_k$ formula is equal to the denominator in the $c_1$ formula (\ref{c1}). Thus, the above $c_k$ formula can be reduced to
\[ \label{ck2}
c_k=-\frac{
\left\langle \left[\begin{array}{c}  \phi^A \\ \psi^A \end{array}\right], \hspace{0.05cm}
\left[\begin{array}{cc} {\cal M}_{11} & {\cal M}_{12} \\
{\cal M}_{21} & {\cal M}_{22} \end{array}\right]\left[\begin{array}{c}
p_{k,s} \\ q_{k,s}\end{array}\right]+\left[\begin{array}{c} {\cal N}_{k+1}^{[1]} \\ {\cal N}_{k+1}^{[2]}\end{array}\right]\right\rangle}
{\int_{-\infty}^\infty \left(\phi^A f_{2b}+\psi^Ag_{2b}\right)\hspace{0.03cm} \hspace{0.04cm} {\rm{d}}x}, \quad k\ge 2.
\]
This process is then repeated to higher orders.

The only conditions for the above perturbation calculations to succeed to all orders are that the numerator and denominator in the $c_0$ formula (\ref{c02}), as well as the denominator in the $c_1$ formula (\ref{c1}), are all nonzero. Thus, we only have 3 numbers to check, which can be easily done for each given equation (\ref{NLS1D}) when its Wadati potential $V(x)$ is specified.

\subsection{Comparison with numerics}
In this subsection, we compare the above perturbation-series soliton solution (\ref{pexpand})-(\ref{qexpand})
with the high-accuracy numerical solution, for a continuous range of small $\epsilon$ values, and confirm the asymptotic accuracy of this analytical solution.

In our comparison, we choose the non-\PT-symmetric Wadati potential (\ref{V}) as the one with
\[ \label{gxform}
g(x)=0.8\left[\mbox{sech}(x+2)+1.2\mbox{sech}(x-2)\right].
\]
The resulting Wadati potential is shown in Fig. 1(a). This potential admits a discrete real eigenvalue $\mu_0\approx 0.37080447$, and its corresponding eigenfunction $\phi(x)+\rm{i}\psi(x)$ is plotted in Fig. 1(b). Numerically, we find that the adjoint operator ${\cal L}^A$ in Eq.~(\ref{calLA}) indeed admits a single localized function $(\phi^A, \psi^A)^T$ in its kernel, and this function is displayed in Fig. 1(c). Utilizing these eigenfunctions and adjoint eigenfunctions, the ratio of integrals under the square root in Eq. (\ref{c02}) is found to be positive. Thus, according to our perturbation theory, a continuous family of solitons would bifurcate out for $\mu>\mu_0$ under the positive sign of nonlinearity $\sigma=1$ and for $\mu<\mu_0$ under the negative sign of nonlinearity $\sigma=-1$.

Numerically, this is found to be the case. With the choice of positive sign of nonlinearity $\sigma=1$, this soliton at $\mu=\mu_0+0.1$ is
exhibited in Fig. 1(d). In addition, the power function of this soliton family, defined as
\[ \label{Pmu}
P(\mu)=\int_{-\infty}^\infty |u(x; \mu)|^2 \hspace{0.04cm} {\rm{d}}x,
\]
is shown in Fig. 1(e). These solitons are computed numerically by the Newton-conjugate-gradient method described in \cite{myPRA2D}, and their numerical error is below $10^{-10}$. Due to their high accuracy, we will call these numerical solutions as exact solutions in the remainder of this subsection.

Now, we make a more quantitative comparison between our perturbation-series solution and the exact solution. For this purpose, we first consider the perturbation-series solution (\ref{pexpand})-(\ref{qexpand}) at $\mu=\mu_0+0.1$, i.e., when $\epsilon=0.1$. This analytical solution, to the third order of the perturbation series, is determined from the formulae and equations for $(c_0, c_1, c_2)$, $(p_0, q_0)$, $(p_1, q_1)$ and $(p_2, q_2)$ in the previous subsection, and plotted in Fig. 1(d) alongside the exact solution. As can be seen, this third-order perturbation solution is almost indistinguishable from the exact solution. This is not surprising, since this third-order perturbation solution has relative error of order $\epsilon^{3}$, or roughly 0.001 for $\epsilon=0.1$, which is indeed very small.

Next, we compare the power function of our perturbation-series solutions (\ref{pexpand})-(\ref{qexpand}) to that of the exact soliton solutions. For this purpose, we insert the perturbation-series solution (\ref{pexpand})-(\ref{qexpand}) into the power function definition (\ref{Pmu}) and get
\[ \label{Pexpand}
P_{anal} \ (\mu)=\epsilon P_1+\epsilon^2 P_2 + \epsilon^3 P_3 + \cdots,
\]
where $\mu=\mu_0+\epsilon$ as before, and
\begin{equation*}
P_1=\int_{-\infty}^\infty (p_0^2+q_0^2) \hspace{0.04cm} {\rm{d}}x, \quad P_2=\int_{-\infty}^\infty 2(p_0p_1+q_0q_1) \hspace{0.04cm} {\rm{d}}x, \quad P_3=\int_{-\infty}^\infty \left[p_1^2+q_1^2+2(p_0p_2+q_0q_2)\right] \hspace{0.04cm} {\rm{d}}x.
\end{equation*}
Using the $(p_0, q_0)$, $(p_1, q_1)$ and $(p_2, q_2)$ solutions we have numerically obtained, we find that
\begin{equation}
P_1\approx 5.89609348, \quad P_2\approx -5.65066426, \quad P_3\approx -9.38398099.
\end{equation}
Truncating the power-function expansion (\ref{Pexpand}) to the third order, this truncated power function is plotted in Fig. 1(e) alongside the exact power function. Again, the two functions are almost indistinguishable when $\mu$ is close to $\mu_0$.

The power series (\ref{Pexpand}) is an asymptotic series. It does not have to be convergent, but it must satisfy the requirement of an asymptotic series, which is that $|P(\mu)-\sum_{k=1}^n \epsilon^k P_k|=o(\epsilon^n)$ when $\epsilon\to 0$ for every positive integer $n$ \cite{BenderOrszag}. To verify this asymptotic condition of our power series (\ref{Pexpand}), we examine the difference between the third-order truncated power expansion (\ref{Pexpand}) and the exact power function. According to our power expansion, this difference is expected to be
\[ \label{deltaP}
\Delta P \equiv P(\mu) - \epsilon P_1 - \epsilon^2 P_2 - \epsilon^3 P_3 = O(\epsilon^4).
\]
If this is indeed true, then the above asymptotic condition for $n=3$ would be met. To confirm this $\Delta P=O(\epsilon^4)$ asymptotics for small $\epsilon$, we show in Fig. 1(f) a log-log plot of $\Delta P$ versus $\epsilon$. Its comparison with the benchmark $\Delta P=\epsilon^4$ curve on the same graph shows that this $\Delta P$ is indeed $O(\epsilon^4)$ at small $\epsilon$, confirming the asymptotic accuracy of our third-order power expansion.

\begin{figure}[htbp]
\begin{center}
\includegraphics[width=0.8\textwidth]{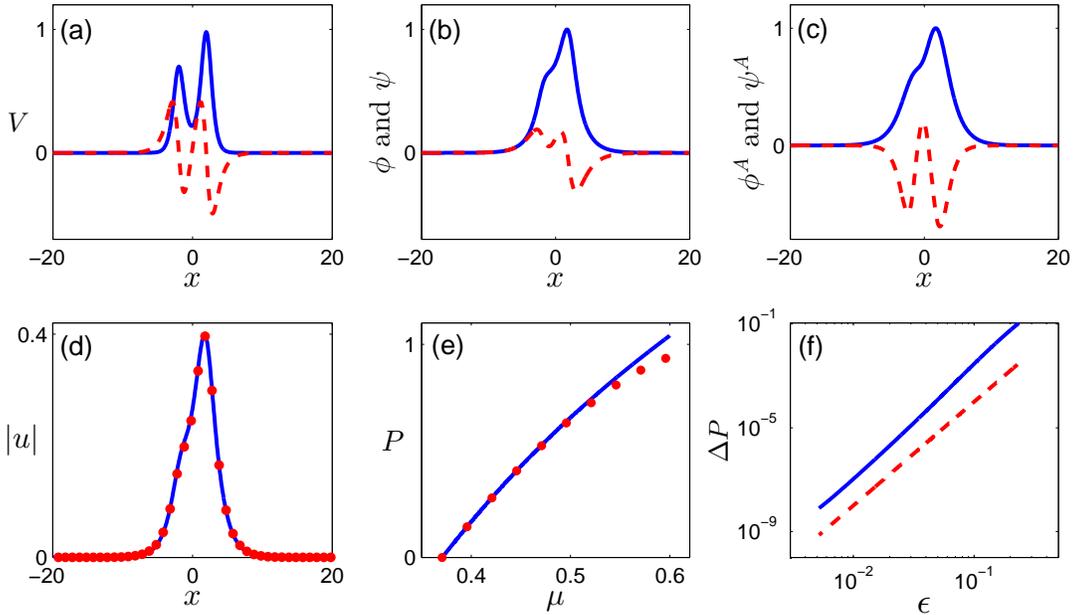}
\caption{Comparison of solitons between theory and numerics for the 1D equation (\ref{ueq}) with $\sigma=1$ and $g(x)$ given by Eq. (\ref{gxform}). (a) Wadati potential (\ref{V}), where solid blue is Re$(V)$ and dashed red Im$(V)$. (b) Linear eigenmode $\phi(x)+\rm{i}\psi(x)$ of this potential, with solid blue being $\phi(x)$ and dashed red $\psi(x)$. (c) Localized adjoint eigenfunction, with solid blue being $\phi^A(x)$ and dashed red $\psi^A(x)$. (d) Amplitude profile $|u(x;\mu)|$ of the soliton at $\mu=\mu_0+0.1$, where solid blue is from numerical computation and red dots from analytical third-order perturbation series prediction. (e) Power curve of this soliton family, with solid blue from numerical computations and red dots from the third-order perturbation expansion (\ref{Pexpand}). (f) Log-log plot of the power difference (\ref{deltaP}) between numerical values and the third-order perturbation expansion versus $\epsilon=\mu-\mu_0$. The dashed red line is the $\Delta P=\epsilon^4$ curve for comparison. }
\end{center}
\end{figure}

In the above numerical example, we chose the focusing nonlinearity (with $\sigma=1$). If the nonlinearity is defocusing, we have found similarly good agreement between perturbation-series solutions and the numerics.

\subsection{An alternative perturbation calculation}

In the above perturbation calculation, we introduced the tangible small parameter as $\epsilon=\mu-\mu_0$. Because of that, we only needed to expand the solutions $(p, q)$ into perturbation series. In this treatment, the $(p_n, q_n)^T$ solution at each order must contain the homogeneous term $c_n (\phi, \psi)^T$, so that $c_n$ can be selected judiciously to satisfy the solvability condition of the linear nonhomogeneous $(p_{n+1}, q_{n+1})^T$ equation.

There is an alternative perturbation calculation, where we expand not only the solutions $(p, q)$, but also the propagation constant $\mu$, into perturbation series. In this treatment, the $(p, q)$ expansion would still be (\ref{pexpand})-(\ref{qexpand}), while the $\mu$ expansion would be
\[
\mu=\mu_0+\mu_1 \epsilon +\mu_2 \epsilon^2 + \cdots,
\]
where $\mu_1, \mu_2, \cdots$ are real constants to be determined. Due to the introduction of these $(\mu_1, \mu_2, \cdots)$ parameters in the $\mu$-expansion, we can choose each $\mu_n$ judiciously to satisfy the solvability condition of the linear nonhomogeneous $(p_{n}, q_{n})^T$ equation. As a consequence, we do not need to introduce the homogeneous term $c_n (\phi, \psi)^T$ in the $(p_{n}, q_{n})^T$ solution anymore. In this alternative treatment, we systematically detune the propagation constant $\mu$; while in the original treatment, we systematically detune the coefficient of the $(\phi, \psi)^T$ term in the $(p, q)^T$ solution, since that coefficient is $c_0+c_1\epsilon+c_2\epsilon^2+\cdots$. Algebra-wise, this alternative perturbation calculation turns out to be a little simpler, because $\mu$ appears in the original two real soliton equations (\ref{peq})-(\ref{qeq}) in a simpler way than $p$ and $q$, and thus this $\mu$-detuning introduces less terms in each $(p_{n}, q_{n})$ equation than our present treatment. The slight downside of this alternative treatment is that, the ``physical" meaning of the small parameter $\epsilon$ in it is less clear. Indeed, $\epsilon$ in this alternative treatment is more like a non-tangible arbitrary book-keeping-type small parameter, to which both the propagation constant $\mu$ and the soliton solution $(p, q)$ relate in a nontrivial parametric (perturbation-series) way. Overall, these two different perturbation procedures are roughly equivalent, and their choice is largely a personal taste. Indeed, we have also implemented this alternative perturbation treatment  analytically and compared its results to the numerics, and found similar agreement as that shown in Fig.~1.

\section{Construction of soliton families in the 2D case}
Now, we consider the 2D NLS equation
\[ \label{NLS2D}
iU_{t}+U_{xx}+U_{yy}+V(x,y)U+\sigma |U|^2U=0,
\]
where $V(x,y)$ is a complex potential, and $\sigma$ the sign of nonlinearity. It has been shown in \cite{NixonYang2016} that when this potential is of the form
\[ \label{Vxy}
V(x,y)=g^2(x)+{\rm{i}}g'(x)+h(y),
\]
where $g(x)$ and $h(y)$ are real functions, then its spectrum can be all-real. This potential is separable, and its $x$-part is the 1D Wadati potential (\ref{V}). So, this 2D potential will also be called Wadati-type in this article. When $g(x)$ is even, then this potential admits the partial \PT symmetry $V^*(x,y)=V(-x, y)$. In this case, Eq. (\ref{NLS2D}) admits continuous families of solitons, which has been demonstrated numerically and explained analytically in \cite{PPT}. However, when $g(x)$ is not even, so that the potential $V(x,y)$ is non-\PT-symmetric, numerical evidence in \cite{myPRA2D} indicates that Eq. (\ref{NLS2D}) could still admit continuous families of solitons, which is mysterious in the absence of \PT symmetry.

In this section, we analytically explain the existence of continuous families of solitons in the 2D NLS equation (\ref{NLS2D}) with a non-\PT-symmetric Wadati-type potential  (\ref{Vxy}) by extending the 1D perturbation calculations of the previous section to the present 2D case. In this potential (\ref{Vxy}), we require $g(x)$ to be localized and differentiable, and $h(y)$ localized and continuous or piece-wise continuous.

Solitons in Eq. (\ref{NLS2D}) are of the form
\[ \label{Uu2D}
U(x,y,t)=e^{{\rm{i}}\mu t}u(x,y),
\]
where $\mu$ is a real propagation constant, and $u(x,y)$ is a localized function satisfying the 2D complex soliton equation
\[ \label{ueq2D}
\left[\partial_{xx}+\partial_{yy}+g^2(x)+{\rm{i}}g'(x)+h(y)\right]u-\mu u +\sigma |u|^2u=0.
\]

\subsection{A real system for 2D solitons and its perturbation expansion}

Similar to the 1D case, an important property of Eq. (\ref{NLS2D}) with the Wadati-type potential (\ref{Vxy}) is that it admits a conservation law even though it is non-Hamiltonian \cite{myPRA2D}. Substituting the soliton solution (\ref{Uu2D}) into that conservation law, we get a stationary real-valued flux equation
\[ \label{flux}
\frac{\partial J_1}{\partial x}+\frac{\partial J_2}{\partial y}=0,
\]
where
\[\label{J1}
J_1=\left|u_x+{\rm{i}}g(x)u\right|^2+\left[h(y)-\mu\right] |u|^2 +\frac{\sigma}{2} |u|^4-|u_y|^2,
\]
and
\[\label{J2}
J_2=u_xu_y^*+u_x^*u_y-{\rm{i}}g(x)(u_yu^*-uu_y^*).
\]
Following the 1D strategy, instead of working with the complex soliton equation (\ref{ueq2D}), we will work with the real part of that soliton equation, i.e.,
\[ \label{peq2D}
\left[\partial_{xx}+\partial_{yy}+g^2(x)+h(y)-\mu\right]p-g'(x)q+\sigma(p^2+q^2)p=0,
\]
where $u\equiv p+{\rm i}q$ as before [see (\ref{upq})], together with the real-valued flux equation (\ref{flux}), in our construction of a continuous family of 2D solitons. A minor difference from the 1D case is that here, we have to use the flux equation (\ref{flux}), which is the counterpart of the $dJ/dx=0$ equation in the 1D case. This contrasts the 1D case where we used $J=0$ directly. This minor difference in the starting equations for solitons will lead to minor differences in the technical constructions of soliton solutions, as we will see later in this section.

The soliton family to be constructed bifurcates from a discrete real eigenvalue $\mu_0$ of the potential. The corresponding localized eigenmode $\hat{\phi}(x,y)+{\rm{i}}\hat{\psi}(x,y)$, with real $(\hat{\phi}, \hat{\psi})$, satisfies the linear eigenmode equation obtained by dropping the nonlinear term in the soliton equation (\ref{ueq2D}), i.e.,
\[ \label{eigcomp2D}
\left[\partial_{xx}+\partial_{yy} +g^2(x)+{\rm{i}}g'(x)+h(y)\right] (\hat{\phi}+{\rm{i}}\hat{\psi})=\mu_0 (\hat{\phi}+{\rm{i}}\hat{\psi}).
\]
Since the potential in this equation is separable, its linear mode is also separable and can be decomposed as
\[ \label{hatphipsi}
\hat{\phi}(x,y)=\phi(x) \zeta(y), \quad \hat{\psi}(x,y)=\psi(x)\zeta(y),
\]
and $\mu_0=\mu_{01}+\mu_{02}$, where $\phi(x)+{\rm{i}}\psi(x)$ is a localized eigenmode of the $x$-part of the potential (a Wadati potential) with a discrete real eigenvalue $\mu_{01}$, i.e.,
\[ \label{eigcompx}
\left[\partial_{xx}+g^2(x)+{\rm{i}}g'(x)\right] (\phi+{\rm{i}}\psi)=\mu_{01} (\phi+{\rm{i}}\psi),
\]
and $\zeta(y)$ is a real localized eigenmode of the $y$-part of the potential with a discrete real eigenvalue $\mu_{02}$, i.e.,
\[ \label{eigcompy}
\left[\partial_{yy}+h(y)\right]\zeta =\mu_{02} \zeta.
\]
This 2D eigenmode $u=\hat{\phi}(x,y)+{\rm{i}}\hat{\psi}(x,y)$ satisfies the flux equation (\ref{flux}) with the $|u|^4$ term dropped in $J_1$ and $\mu$ replaced by $\mu_0$.

Bifurcating from this linear eigenmode, we seek a low-amplitude soliton at each real propagation constant value $\mu$ near the linear eigenvalue $\mu_0$, and this soliton is expanded into the following perturbation series,
\begin{eqnarray}
&& p(x,y; \mu) = \epsilon^{1/2}\left[ p_0(x,y)+\epsilon p_1(x,y) +\epsilon^2 p_2(x,y) +\cdots\right],   \label{pexpand2D}\\
&& q(x,y; \mu) = \epsilon^{1/2}\left[ q_0(x,y)+\epsilon q_1(x,y) +\epsilon^2 q_2(x,y) +\cdots\right],   \label{qexpand2D}
\end{eqnarray}
where $\epsilon=\mu-\mu_0$ and is assumed to be small positive (so that $\epsilon^{1/2}$ is real). As explained in the 1D case, this positive-$\epsilon$ assumption corresponds to a proper sign of nonlinearity $\sigma$, and the negative-$\epsilon$ case can be treated similarly.

Substituting the above perturbation expansion into Eqs. (\ref{flux}) and (\ref{peq2D}), we get a sequence of real equations for $(p_k, q_k)$. The equations for $(p_0, q_0)$ are just the flux equation (\ref{flux}) with the $|u|^4$ term dropped in $J_1$, and the linear part of Eq. (\ref{peq2D}), with $\mu$ replaced by $\mu_0$. Their solutions are obviously
\[ \label{p0q02D}
\left[\begin{array}{c} p_0 \\ q_0 \end{array}\right]=c_0 \left[\begin{array}{c} \hat{\phi} \\ \hat{\psi} \end{array}\right],
\]
where $c_0$ is a real constant to be determined. The equations for $(p_k, q_k)$ ($k\ge 1$) are the following linear nonhomogeneous system of equations
\[ \label{Lpkqk2D}
\widehat{\cal L}\left[\begin{array}{c} p_k \\ q_k \end{array}\right]= \left[\begin{array}{c} \hat{f}_k \\
\partial_x\hat{g}_{k1}+\partial_y\hat{g}_{k2} \end{array}\right],
\]
where $\widehat{\cal L}$ is a $2\times 2$ matrix operator whose elements are
\begin{eqnarray*}
&& \widehat{\cal L}_{11}=\partial_{xx}+\partial_{yy}+g^2+h-\mu_0, \\
&& \widehat{\cal L}_{12}=-g_x, \\
&& \widehat{\cal L}_{21}=\partial_x \left[(\hat{\phi}_x-g\hat{\psi})\partial_x+g(\hat{\psi}_x+g\hat{\phi})+(h-\mu_0)\hat{\phi}-\hat{\phi}_y\partial_y\right]+\partial_y \left[
\hat{\phi}_y\partial_x+(\hat{\phi}_x-g\hat{\psi})\partial_y + g\hat{\psi}_y\right], \\
&& \widehat{\cal L}_{22}=\partial_x\left[(\hat{\psi}_x+g\hat{\phi})\partial_x-g(\hat{\phi}_x-g\hat{\psi})+(h-\mu_0)\hat{\psi}-\hat{\psi}_y\partial_y\right]+
\partial_y \left[\hat{\psi}_y\partial_x+(\hat{\psi}_x+g\hat{\phi})\partial_y -g\hat{\phi}_y \right],
\end{eqnarray*}
\[ \label{f1g12D}
\left[\begin{array}{c} \hat{f}_1 \\ \hat{g}_{11} \\  \hat{g}_{12}\end{array}\right]= c_0 \left[\begin{array}{l} \hat{\phi} -\sigma c_0^2 (\hat{\phi}^2+\hat{\psi}^2)\hat{\phi} \\
\frac{1}{2}(\hat{\phi}^2+\hat{\psi}^2)-\frac{1}{4}\sigma c_0^2 (\hat{\phi}^2+\hat{\psi}^2)^2 \\ 0\end{array}\right],
\]
\[ \label{f2g22D}
\left[\begin{array}{c} \hat{f}_2 \\ \hat{g}_{21} \\  \hat{g}_{22} \end{array}\right]= \left[\begin{array}{l} \left(1 -3\sigma p_0^2 -\sigma q_0^2\right)p_1-2\sigma p_0q_0q_1 \\ \frac{1}{2c_0}\left[2\left(1-\sigma p_0^2-\sigma q_0^2\right)(p_0p_1+q_0q_1)+(\mu_0-h)(p_1^2+q_1^2)+(p_{1y}^2+q_{1y}^2)-(p_{1x}-gq_1)^2-(q_{1x}+gp_1)^2)\right] \\
-\frac{1}{c_0}\left[(p_{1x}-gq_1)p_{1y}+(q_{1x}+gp_1)q_{1y}\right]
\end{array}\right],
\]
\[ \label{fkgk2D}
\left[\begin{array}{c} \hat{f}_k \\ \hat{g}_{k1} \\  \hat{g}_{k2} \end{array}\right]= \left[\begin{array}{cc} \widehat{{\cal M}}_{11} & \widehat{{\cal M}}_{12} \\
\widehat{{\cal M}}_{21} & \widehat{{\cal M}}_{22} \\ \widehat{{\cal M}}_{31} & \widehat{{\cal M}}_{32}
\end{array}\right]\left[\begin{array}{c}
p_{k-1} \\ q_{k-1}\end{array}\right] +\left[\begin{array}{c} \widehat{{\cal N}}_{k}^{[1]} \\ \widehat{{\cal N}}_{k}^{[2]}\end{array}\right],    \quad k\ge 3,
\]
the matrix elements $\widehat{{\cal M}}_{ij}$ are $k$-independent and given by the formulae
\begin{eqnarray*}
&& \widehat{{\cal M}}_{11}=1-3\sigma p_0^2-\sigma q_0^2, \\
&& \widehat{{\cal M}}_{12}=-2\sigma p_0q_0, \\
&& \widehat{{\cal M}}_{21}=\frac{1}{c_0}\left[p_0\left( 1-\sigma p_0^2-\sigma q_0^2\right)+(\mu_0-h)p_1+p_{1y}\partial_y-(p_{1x}-gq_1)\partial_x-g(q_{1x}+gp_1)\right],\\
&& \widehat{{\cal M}}_{22}=\frac{1}{c_0}\left[q_0\left( 1-\sigma p_0^2-\sigma q_0^2\right)+(\mu_0-h)q_1
+q_{1y}\partial_y+g(p_{1x}-gq_1)-(q_{1x}+gp_1)\partial_x\right], \\
&& \widehat{{\cal M}}_{31}=(p_{1x}-gq_1)\partial_y+p_{1y}\partial_x+gq_{1y}, \\
&& \widehat{{\cal M}}_{32}=(q_{1x}+gp_1)\partial_y+q_{1y}\partial_x-gp_{1y},
\end{eqnarray*}
and $\widehat{{\cal N}}_{k}^{[1]}, \widehat{{\cal N}}_{k}^{[2]}$ are functions which depend only on $k$, $p_0, p_1, \dots, p_{k-2}$, $q_0, q_1, \dots, q_{k-2}$,  $g(x)$ and $h(y)$.

\subsection{Kernel structures of the 2D linear operator and its adjoint operator}

To solve the 2D linear nonhomogeneous equations (\ref{Lpkqk2D}) and obtain localized solutions $(p_k, q_k)$ for all $k$, we will also use the Fredholm alternative method. To do so, we need to understand the kernel structures of the 2D operator $\widehat{\cal L}$ and its adjoint operator $\widehat{\cal L}^A$, where elements of the adjoint operator are
\begin{eqnarray*}
&& \widehat{\cal L}^A_{11}=\partial_{xx}+\partial_{yy}+g^2+h-\mu_0, \\
&& \widehat{\cal L}^A_{21}=-g_x, \\
&& \widehat{\cal L}^A_{12}= \left[\partial_x(\hat{\phi}_x-g\hat{\psi})-g(\hat{\psi}_x+g\hat{\phi})-(h-\mu_0)\hat{\phi}-\partial_y\hat{\phi}_y\right]\partial_x+
\left[\partial_x\hat{\phi}_y+\partial_y(\hat{\phi}_x-g\hat{\psi}) - g\hat{\psi}_y\right]\partial_y,  \\
&& \widehat{\cal L}^A_{22}=\left[\partial_x(\hat{\psi}_x+g\hat{\phi})+g(\hat{\phi}_x-g\hat{\psi})-(h-\mu_0)\hat{\psi}-\partial_y\hat{\psi}_y\right]\partial_x+
 \left[\partial_x\hat{\psi}_y+\partial_y(\hat{\psi}_x+g\hat{\phi}) +g\hat{\phi}_y \right]\partial_y.
\end{eqnarray*}

First, we consider the kernel structure of $\widehat{\cal L}$. It is easy to check that this kernel contains two localized functions
\[ \label{K1K22D}
\widehat{K}_1\equiv \left[\begin{array}{c} \hat{\phi} \\ \hat{\psi} \end{array}\right], \quad
\widehat{K}_2\equiv \left[\begin{array}{c} -\hat{\psi} \\ \hat{\phi} \end{array}\right],
\]
where
\[
\widehat{\cal L}\widehat{K}_1=\widehat{\cal L}\widehat{K}_2=0,
\]
similar to the 1D case and for similar reasons. Since the kernel equation $\widehat{\cal L}\widehat{K}=0$ is the linearization of the two real ``eigenvalue" equations for $(p_0, q_0)$ [the 2D counterparts of 1D equations (\ref{p0q01})-(\ref{p0q02})] around the linear mode $(\hat{\phi}, \hat{\psi})$, localized functions in $\widehat{\cal L}$'s kernel can only be induced by amplitude and phase invariances of these $(p_0, q_0)$ equations, which result in $\widehat{K}_1$ and $\widehat{K}_2$ above. Thus, there are no other localized functions in $\widehat{\cal L}$'s kernel.

Next, we consider the kernel structure of the adjoint 2D operator $\widehat{\cal L}^A$. Due to the separability of the 2D eigenmode $(\hat{\phi}, \hat{\psi})$ in Eq. (\ref{hatphipsi}), we can quickly verify that the kernel of $\widehat{\cal L}^A$ contains a bounded function
\[ \label{K0A2D}
\widehat{K}_0^A=\left[\begin{array}{c} \phi^A(x)\hspace{0.05cm} \zeta(y) \\ -\int \psi^A(x) {\rm{d}}x \end{array}\right],
\]
where $\widehat{\cal L}^A\widehat{K}_0^A=0$, and $[\phi^A(x), \psi^A(x)]^T$ is the unique localized function (\ref{K0A}) in the kernel of the 1D adjoint operator ${\cal L}^A$ given in Eq. (\ref{calLA}), with $\mu_0$ replaced by $\mu_{01}$. One may notice that this kernel function of the 2D adjoint operator does not naturally fall back to the 1D adjoint kernel function (\ref{K0A}). The reason is twofold. One is that the second column of the 2D adjoint operator $\widehat{\cal L}^A$, i.e., $[\widehat{\cal L}^A_{12}, \widehat{\cal L}^A_{22}]^T$ given above, contains an additional spatial derivative compared to the second column of the 1D adjoint operator ${\cal L}^A$ given in Eq.~(\ref{calLA}) --- a difference caused by our using the divergence form of the flux equation (\ref{flux}) in 2D instead of its integrated form $J(x)=0$ in 1D. This difference in the second column of the adjoint operator explains the integral in the second element of $\widehat{K}_0^A$ above.
The second reason for $\widehat{K}_0^A$ in 2D not naturally falling back to ${K}_0^A$ in 1D is that, the second columns of the two adjoint operators contain linear eigenmodes or their derivatives as multiplicative factors, while the first columns of these adjoint operators do not. Thus, in the 2D case, we need to introduce the factor $\zeta(y)$ from the 2D linear eigenmode (\ref{hatphipsi}) into the first element of the adjoint kernel function $\widehat{K}_0^A$ in Eq.~(\ref{K0A2D}) in order to balance such a term coming from the second column of $\widehat{\cal L}^A$.

We can further show that, if $h(y)$ is a slowly varying function, then the above $\widehat{K}_0^A$ would be the only bounded function in the kernel of $\widehat{\cal L}^A$. To do so, let $h(y)=\hat{\epsilon}^2H(Y)$ be a slowly varying function of $Y=\hat{\epsilon}y$, where $\hat{\epsilon}$ is a small real parameter. For this $h(y)$, its eigenmode from Eq. (\ref{eigcompy}) is $\zeta(y)=\hat{\zeta}(Y)$, with eigenvalue $\mu_{02}=O(\hat{\epsilon}^2)$.
In this case, $\widehat{\cal L}^A$ can be rewritten as a quadratic function of $\hat{\epsilon}$,
\[ \label{hatLA}
\widehat{\cal L}^A=\widehat{\cal L}^A_0(x,Y)+\hat{\epsilon}\widehat{\cal L}^A_1(x,Y)+\hat{\epsilon}^2\widehat{\cal L}^A_2(x, Y),
\]
where
\begin{eqnarray}
&&\widehat{\cal L}^A_0(x,Y) =\left[\begin{array}{ll} \partial_{xx}+g^2-\mu_{01}  & \hat{\zeta}(Y)\left[
-\partial_x(\phi'-g\psi)+g(\psi'+g\phi)-\mu_{01}\phi\right](-\partial_x) \\
-g'  & \hat{\zeta}(Y)\left[-\partial_x(\psi'+g\phi)-g(\phi'-g\psi)-\mu_{01}\psi\right](-\partial_x) \end{array}\right] \nonumber \\
&&\hspace{1.3cm} ={\cal L}^A \left[\begin{array}{cc} 1 & 0 \\ 0 & -\hat{\zeta}(Y)\partial_x \end{array}\right],
\end{eqnarray}
and ${\cal L}^A$ is the 1D adjoint operator (\ref{calLA}) with $\mu_0$ replaced by $\mu_{01}$.
Since $\widehat{\cal L}^A$ is a function of $x$, $Y$ and $\hat{\epsilon}$, functions $\widehat{F}$ in its kernel are also functions of these same variables and can be expanded into a perturbation series of $\hat{\epsilon}$ as
\[
\widehat{F}(x, y; \hat{\epsilon})=\widehat{F}_0(x,Y)+\hat{\epsilon}\widehat{F}_1(x,Y)+\hat{\epsilon}^2\widehat{F}_2(x, Y)+\dots.
\]
Inserting this expansion and Eq.~(\ref{hatLA}) into $\widehat{\cal L}^A\widehat{F}=0$ and using the kernel structures of the 1D operators  ${\cal L}$ and its adjoint ${\cal L}^A$ detailed in Sec. \ref{sec_Kernel}, we can sequentially determine $\widehat{F}_n(x,Y)$ in the above perturbation expansion and show that the only bounded function in the kernel of $\widehat{\cal L}^A$ is
\[
\widehat{F}=\left[\begin{array}{c} \phi^A(x)\hspace{0.05cm} \hat{\zeta}(Y) \\ -\int \psi^A(x) \hspace{0.05cm} {\rm{d}}x \end{array}\right],
\]
which matches (\ref{K0A2D}) when the eigenmode $\zeta(y)=\hat{\zeta}(Y)$ is slowly varying. All other functions in the kernel of $\widehat{\cal L}^A$ grow exponentially at large $|x|$ or $|Y|$.

When $h(y)$ continuously deforms from slowly varying to the general case of non-slowly varying, the above kernel structure of $\widehat{\cal L}^A$ generically will not change, i.e., its kernel will generically still contain a single bounded function (\ref{K0A2D}). While we cannot at this time rule out the possibility of additional bounded functions appearing in the kernel of $\widehat{\cal L}^A$ at some special $h(y)$ functions during this deformation process, for specific examples of the potentials, we can use numerics to directly verify this single-bounded-function kernel structure for $\widehat{\cal L}^A$, so that our analysis below can proceed.

\subsection{Construction of perturbation series to all orders in 2D} \label{sec:pert2D}
With the above kernel structures of $\widehat{\cal L}$ and $\widehat{\cal L}^A$ in hand, we can now sequentially solve Eq. (\ref{Lpkqk2D}) for localized solutions $(p_k, q_k)$ using the Fredholm alternative method. According to this method, if functions $(\hat{f}_k, \hat{g}_{k1}, \hat{g}_{k2})$ on the right side of the linear nonhomogeneous system (\ref{Lpkqk2D}) are localized (which is the case here), this system would admit a localized solution $(p_k, q_k)$ if and only if its right hand side is orthogonal to the bounded function $\widehat{K}_0^A$ of (\ref{K0A2D}) in the kernel of $\widehat{\cal L}^A$, i.e.,
\[ \label{Fredcond2D}
\left\langle \left[\begin{array}{c}  \phi^A(x)\hspace{0.05cm} \zeta(y) \\  -\int \psi^A(x) \hspace{0.05cm} {\rm{d}}x \end{array}\right], \hspace{0.05cm}
\left[\begin{array}{c}  \hat{f}_k \\ \partial_x\hat{g}_{k1}+\partial_y\hat{g}_{k2} \end{array}\right]\right\rangle=0.
\]
It is noted that the arbitrary constant out of the indefinite integral $\int \psi^A(x) \hspace{0.05cm} {\rm{d}}x$ gives no contribution to the inner product in the above solvability condition. In addition, the above integral is convergent since $(\hat{f}_k, \hat{g}_{k1}, \hat{g}_{k2})$ are all localized in space.

Our perturbative construction of 2D solitons bifurcating from a linear localized eigenmode of the complex potential (\ref{Vxy}) proceeds similarly as the 1D case, since the kernel structures in the 2D case resemble those in the 1D case. We first consider Eq. (\ref{Lpkqk2D}) for $(p_1, q_1)$. Substituting the $(\hat{f}_1, \hat{g}_{11}, \hat{g}_{12})$ expressions (\ref{f1g12D}) into the above solvability condition and performing integration by parts, we get
\[
\left\langle \left[\begin{array}{c}  \phi^A(x)\hspace{0.05cm} \zeta(y) \\  \psi^A(x) \end{array}\right], \hspace{0.05cm}
\left[\begin{array}{c}  \hat{f}_1 \\ \hat{g}_{11} \end{array}\right]\right\rangle=0.
\]
Inserting the $(\hat{f}_1, \hat{g}_{11})$ expressions (\ref{f1g12D}) and $(\hat{\phi}, \hat{\psi})$ formulae (\ref{hatphipsi}) into the above equation, we obtain a formula for $c_0$ as
\[ \label{c022D}
c_0 =\pm \sqrt{ \frac{\int_{-\infty}^\infty \left[\phi \phi^A+\frac{1}{2}(\phi^2+\psi^2)\psi^A\right] \hspace{0.04cm} {\rm{d}}x}{\sigma \int_{-\infty}^\infty \left[ (\phi^2+\psi^2)\phi\phi^A + \frac{1}{4} (\phi^2+\psi^2)^2\psi^A\right] \hspace{0.04cm} {\rm{d}}x} \hspace{0.1cm} \frac{\int_{-\infty}^\infty\zeta^2(y)\hspace{0.04cm} {\rm{d}}y}{\int_{-\infty}^\infty\zeta^4(y)\hspace{0.04cm} {\rm{d}}y}}.
\]
As in the 1D case, the sign of $\sigma$ must match the sign of the ratio between integrals in the above equation so that the quantity under the square root is positive. In addition, we can choose the plus sign outside the square root without loss of generality.

When $c_0$ is selected from the above formula (\ref{c022D}), Eq. (\ref{Lpkqk2D}) admits a localized solution for $(p_1, q_1)$, which we denote as $(p_{1s}, q_{1s})$. Since the kernel of the homogeneous operator $\widehat{\cal L}$ in Eq. (\ref{Lpkqk2D}) contains two localized functions $\widehat{K}_1$ and $\widehat{K}_2$ given in Eq. (\ref{K1K22D}), the general localized solution $(p_1, q_1)$ to the linear nonhomogeneous equations (\ref{Lpkqk2D}) is then $[p_{1s}, q_{1s}]^T+c_1 \widehat{K}_1+d_1\widehat{K}_2$, where $c_1$ and $d_1$ are two real constants. But as in the 1D case, the $d_1$ term can be removed by phase invariance of the complex soliton solution $u(x,y)$. Thus, the $(p_1, q_1)$ solution can be set as
\[ \label{p1q122D}
\left[\begin{array}{c}  p_1 \\ q_1 \end{array}\right]=\left[\begin{array}{c}  p_{1s} \\ q_{1s} \end{array}\right]+ c_1 \left[\begin{array}{c} \hat{\phi} \\ \hat{\psi} \end{array}\right].
\]
The constant $c_1$ in this solution will be determined from the Fredholm solvability condition on the $(p_2, q_2)$ equations.

The equations for $(p_2, q_2)$ are (\ref{Lpkqk2D}), where $(\hat{f}_2, \hat{g}_{21}, \hat{g}_{22})$ in the nonhomogeneous terms are given in Eq. (\ref{f2g22D}). Substituting the $(p_0, q_0)$ solutions (\ref{p0q02D}) and $(p_1, q_1)$ solutions (\ref{p1q122D}) into these nonhomogeneous terms and recalling that the eigenmode $u=\hat{\phi}(x,y)+{\rm{i}}\hat{\psi}(x,y)$ satisfies the flux equation (\ref{flux}) with the $|u|^4$ term dropped in $J_1$ and $\mu$ replaced by $\mu_0$, we see that the right side of Eq. (\ref{Lpkqk2D}) for $(p_2, q_2)$ reduces to
\[ \label{f2g2b2D}
\left[\begin{array}{c} \hat{f}_2 \\ \partial_x\hat{g}_{21}+\partial_y\hat{g}_{22} \end{array}\right]=\left[\begin{array}{c} \hat{f}_{2a} \\ \partial_x\hat{g}_{21a}+\partial_y\hat{g}_{22a} \end{array}\right]+c_1 \left[\begin{array}{c} \hat{f}_{2b} \\  \partial_x\hat{g}_{21b}+\partial_y\hat{g}_{22b}\end{array}\right],
\]
where
\begin{eqnarray*}
&&\hat{f}_{2a}=\left(1 -3\sigma p_0^2 -\sigma q_0^2\right)p_{1s}-2\sigma p_0q_0q_{1s}, \\
&&\hat{g}_{21a} = \frac{1}{2c_0}\left[2\left(1-\sigma p_0^2-\sigma q_0^2\right)(p_0p_{1s}+q_0q_{1s})+(\mu_0-h)(p_{1s}^2+q_{1s}^2)+(p_{1s,y}^2+q_{1s,y}^2)  \right. \\
&& \hspace{1.7cm} \left. -(p_{1s,x}-gq_{1s})^2-(q_{1s,x}+gp_{1s})^2)\right], \\
&&\hat{g}_{22a}=-\frac{1}{c_0}\left[(p_{1s,x}-gq_{1s})p_{1s,y}+(q_{1s,x}+gp_{1s})q_{1s,y}\right],
\end{eqnarray*}
and
\begin{eqnarray*}
&&\hat{f}_{2b}= (1-3\sigma p_0^2-\sigma q_0^2)\hat{\phi}-2\sigma p_0q_0\hat{\psi},\\
&&\hat{g}_{21b} = \frac{1}{c_0}\left[\left(1-\sigma p_0^2-\sigma q_0^2\right)(p_0\hat{\phi}+q_0\hat{\psi})+(\mu_0-h)(p_{1s}\hat{\phi}+q_{1s}\hat{\psi})+(p_{1s,y}\hat{\phi}_y+q_{1s,y}\hat{\psi}_y) \right. \\
&& \hspace{1.6cm} \left. -(p_{1s,x}-gq_{1s})(\hat{\phi}_x-g\hat{\psi})-(q_{1s,x}+gp_{1s})(\hat{\psi}_x+g\hat{\phi})\right],  \\
&&\hat{g}_{22b} =-\frac{1}{c_0}\left[(p_{1s,x}-gq_{1s})\hat{\phi}_y+p_{1s,y}(\hat{\phi}_x-g\hat{\psi})
+(q_{1s,x}+gp_{1s})\hat{\psi}_y+q_{1s,y}(\hat{\psi}_x+g\hat{\phi})\right].
\end{eqnarray*}
Inserting (\ref{f2g2b2D}) into the Fredholm solvability condition (\ref{Fredcond2D}) at $k=2$, we get a formula for the constant $c_1$ as
\[ \label{c12D}
c_1=-\frac{\left\langle \left[\begin{array}{c}  \phi^A(x)\hspace{0.05cm} \zeta(y) \\  -\int \psi^A(x) {\rm{d}}x \end{array}\right], \hspace{0.05cm}\left[\begin{array}{c}  \hat{f}_{2a} \\ \partial_x\hat{g}_{21a}+\partial_y\hat{g}_{22a} \end{array}\right]\right\rangle}
{\left\langle \left[\begin{array}{c}  \phi^A(x)\hspace{0.05cm} \zeta(y) \\  -\int \psi^A(x) {\rm{d}}x \end{array}\right], \hspace{0.05cm}\left[\begin{array}{c}  \hat{f}_{2b} \\ \partial_x\hat{g}_{21b}+\partial_y\hat{g}_{22b} \end{array}\right]\right\rangle}.
\]

When the $c_1$ value is selected as above, the $(p_1, q_1)$ solutions (\ref{p1q122D}) are completely determined. In addition, the Fredholm solvability condition (\ref{Fredcond2D}) for the $(p_2, q_2)$ equations (\ref{Lpkqk2D}) is also satisfied; so these equations admit a localized $(p_2, q_2)$ solution, which we denote as $(p_{2s}, q_{2s})$. In view of the kernel structure of operator $\widehat{\cal L}$ and phase invariance of the complex soliton solution $u(x,y)$, the general localized solutions for $(p_2, q_2)$ can be written as
\[ \label{p2q22D}
\left[\begin{array}{c}  p_2 \\ q_2 \end{array}\right]=\left[\begin{array}{c}  p_{2s} \\ q_{2s} \end{array}\right]+ c_2 \left[\begin{array}{c} \hat{\phi} \\ \hat{\psi} \end{array}\right],
\]
where $c_2$ is a real constant. This constant $c_2$ will be determined from the Fredholm solvability condition for the $(p_3, q_3)$ equations. Indeed, inserting this $(p_2, q_2)$ solution into the right side of Eq. (\ref{Lpkqk2D}) with $k=3$, it is easy to see that the solvability condition (\ref{Fredcond2D}) at $k=3$ is a linear equation for $c_2$, which we can easily solve to obtain the value of $c_2$. After this $c_2$ value is obtained, $(p_2, q_2)$ is ascertained. In addition, the $(p_3, q_3)$ equation admits a localized solution, which we denote as $(p_{3s}, q_{3s})$, and the general $(p_3, q_3)$ solutions can be written as (\ref{p2q22D}) with the index changed from 2 to 3. This process is then repeated to higher orders.

\subsection{Comparison with numerics in 2D}
Lastly, we compare the above 2D perturbation series soliton solution (\ref{pexpand2D})-(\ref{qexpand2D}) with the high-accuracy numerical solution and confirm the asymptotic accuracy of this 2D analytical solution. In our comparison, we choose the potential (\ref{Vxy}) with
\[ \label{gxform2D}
g(x)=0.8\left[\mbox{sech}(x+2)+1.2\mbox{sech}(x-2)\right], \quad h(y)=2\mbox{sech}^2y.
\]
Notice that this $g(x)$ function is the same as (\ref{gxform}) in the 1D example. This potential admits a discrete real eigenvalue $\mu_0=\mu_{01}+\mu_{02}\approx 1.37080447$, where $\mu_{01}\approx 0.37080447$ as in the 1D case, and $\mu_{02}=1$. The corresponding eigenfunction $(\hat{\phi}, \hat{\psi})$ is given in Eq. (\ref{hatphipsi}), where $[\phi(x), \psi(x)]$ is as shown in Fig. 1(b), and $\zeta(y)=\mbox{sech}(y)$. Numerically, we confirmed that the 2D adjoint operator $\widehat{\cal L}^A$ indeed admits a single bounded function
(\ref{K0A2D}) in its kernel, where $[\phi^A(x), \psi^A(x)]$ is the localized function (\ref{K0A}) in the kernel of the 1D adjoint operator ${\cal L}^A$ in Eq. (\ref{calLA}), which was plotted in Fig. 2(c).

When $\sigma=1$ (focusing nonlinearity), our theory predicts that a continuous family of solitons bifurcates out from the above linear discrete eigenmode when $\mu>\mu_0$. This is indeed the case. For the choice of $\mu=\mu_0+0.1$ (i.e., $\epsilon=0.1$), the second-order perturbation-series solution (\ref{pexpand2D})-(\ref{qexpand2D}) is determined from the formulae and equations for $c_0, c_1$, $p_0, q_0$,  $p_1$ and $q_1$ in the previous subsection, and plotted in Fig. 2(a). The high-accuracy numerical solution at this same $\mu$ value is displayed in Fig. 2(b) for comparison. It is seen that these two solutions are visually identical. We have also calculated the difference between these two solutions, and found that the relative error between them is under $2.5\%$, which is $O(\epsilon^2)$ (i.e., order of 0.01) as expected.

Next, we compare the power function of our perturbation-series solutions (\ref{pexpand2D})-(\ref{qexpand2D}) to that of the exact soliton solutions. This 2D power function is defined as
\[ \label{Pmu2D}
P(\mu)=\int_{-\infty}^\infty \int_{-\infty}^\infty|u(x,y; \mu)|^2 \hspace{0.04cm} {\rm{d}}x{\rm{d}}y,
\]
analogous to the 1D case (\ref{Pmu}). Inserting the perturbation-series solution (\ref{pexpand2D})-(\ref{qexpand2D}) into this power function, we get
\[ \label{Pexpand2D}
P_{anal} \ (\mu)=\epsilon P_1+\epsilon^2 P_2 +  \cdots,
\]
where
\begin{eqnarray}
&& P_1=\int_{-\infty}^\infty \int_{-\infty}^\infty (p_0^2+q_0^2) \hspace{0.04cm} {\rm{d}}x{\rm{d}}y, \quad
P_2=\int_{-\infty}^\infty \int_{-\infty}^\infty 2(p_0p_1+q_0q_1) \hspace{0.04cm} {\rm{d}}x{\rm{d}}y.
\end{eqnarray}
Using the $(p_0, q_0)$ and $(p_1, q_1)$ solutions obtained from Eqs. (\ref{p0q02D}) and (\ref{p1q122D}), we find that
\begin{equation*}
P_1\approx 17.68828045, \quad P_2\approx -21.74575.
\end{equation*}
Truncating the power-function expansion (\ref{Pexpand2D}) to these first two terms, this truncated power function is plotted in Fig. 2(c) alongside the exact power function. Again, the two functions are almost indistinguishable. To verify the asymptotic accuracy of our perturbation series solutions, we show in Fig. 2(d) a log-log plot of $\Delta P \equiv \epsilon P_1+\epsilon^2 P_2 -P(\mu)$ versus $\epsilon$. Its comparison with the benchmark $\Delta P=\epsilon^3$ curve on the same graph shows that this $\Delta P$ is $O(\epsilon^3)$, which matches our asymptotic prediction for this quantity. The above comparison indicates that the true 2D soliton solutions and our perturbation series solutions are in perfect agreement.

\begin{figure}[htbp]
\begin{center}
\includegraphics[width=0.6\textwidth]{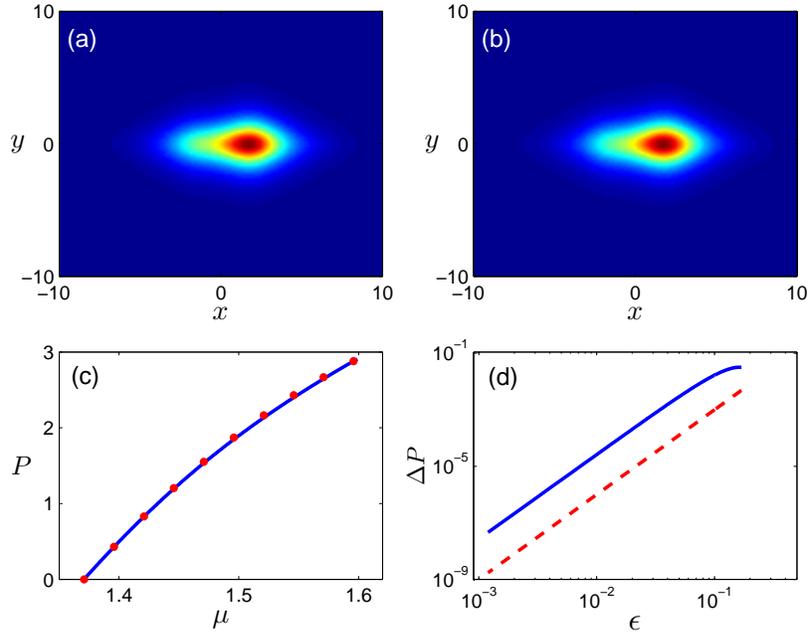}
\caption{Comparison of solitons between theory and numerics for the 2D equation (\ref{ueq2D}) with $\sigma=1$, and $g(x)$, $h(y)$ given by Eq. (\ref{gxform2D}). (a) Amplitude profile $|u(x,y;\mu)|$ of the second-order perturbation series solution (\ref{Pexpand2D}) at
$\mu=\mu_0+0.1$. (b) Numerically computed soliton $|u|$ at the same $\mu$ value of (a). (c) Power curve of this soliton family, with solid blue from numerical computations and red dots from the second-order perturbation expansion (\ref{Pexpand2D}). (d) Log-log plot of the power difference between numerical values and the second-order perturbation expansion versus $\epsilon=\mu-\mu_0$. Dashed red line is the $\Delta P=\epsilon^3$ curve for comparison. }
\end{center}
\end{figure}

\section{Summary and Discussion}

In this article, we have analytically constructed continuous families of low-amplitude solitons bifurcating from linear modes in one- and two-dimensional NLS equations with localized Wadati-type non-\PT-symmetric complex potentials, thus providing an analytical explanation for this counter-intuitive phenomenon of soliton families appearing in these non-\PT-symmetric non-Hamiltonian systems. Our analytical construction utilized the conservation laws of these non-\PT-symmetric equations, which allowed us to convert the complex soliton equations into new real systems. A key advantage of these new real systems is that, during a perturbation expansion of low-amplitude solitons bifurcating from linear modes, the underlying linear operator has two localized functions in its kernel, and the associated adjoint operator has a single localized or bounded function in its kernel. This kernel structure, coupled with the phase invariance of the complex soliton, guarantees that at each order of the soliton's perturbation expansion, the Fredholm solvability condition can always be satisfied, so that a localized solution at each order of the perturbation series can be found. As a result, a continuous family of low-amplitude solitons bifurcating from a linear mode is obtained as a perturbation series to all orders of the small soliton amplitude. We have also compared these analytically constructed soliton solutions to high-accuracy numerical solutions, in both one and two spatial dimensions, and the asymptotic accuracy of these perturbation solutions is fully confirmed.

In this article, the nonlinearity in our 1D and 2D NLS equations (\ref{NLS1D}) and (\ref{NLS2D}) is cubic. But our analytical treatment for this cubic nonlinearity can be trivially generalized to other types of nonlinearities of the general form $G(|U|^2)U$, where $G(\cdot)$ is an arbitrary real function. Indeed, for the 1D and 2D NLS equations (\ref{NLS1D}) and (\ref{NLS2D}) with this more general form of nonlinearity but the same Wadati-type complex potentials (\ref{V}) and (\ref{Vxy}), a conservation law still exists \cite{NixonYangSAMP,myPRA2D}. Thus, the analytical treatment of this article still applies.

In our perturbative construction of soliton families in the NLS equations (\ref{NLS1D}) and (\ref{NLS2D}) with non-\PT-symmetric Wadati-type potentials, the conservation laws of those equations played a critical role. If such conservation laws are absent, such as for non-\PT-symmetric complex potentials not of Wadati-type, this construction would not work. In such cases, we do not believe true soliton families can still exist. This implies that we do not think the ``soliton families" reported in \cite{Panos2019} for non-Wadati complex potentials are true soliton solutions.

A closely related subject is symmetry breaking of solitons in \PT-symmetric Wadati-type potentials (\ref{V}) and (\ref{Vxy}), where $g(x)$ is an even function. It is known that for generic \PT-symmetric potentials, symmetry breaking of solitons is forbidden.
%\cite{YangSAM2014}.
However, for \PT-symmetric Wadati-type potentials (\ref{V}) and (\ref{Vxy}), symmetry breaking of solitons can occur, where two branches of non-\PT-symmetric solitons bifurcate out from the base branch of \PT-symmetric solitons when the base branch's power reaches a certain threshold \cite{myPRA2D,YangOL2014}. So far, there has been no analytical explanation for these symmetry breakings. For \PT-symmetric Wadati-type potentials, the conservation laws (\ref{QJ}) and (\ref{flux}) are still valid. Then, using our new real system of soliton equations in this article, together with bifurcation conditions for symmetry breaking,
%akin to that for the conservative NLS-type equations developed in \cite{YangSAM_classification},
branches of symmetric and asymmetric solitons in these \PT-symmetric Wadati-type potentials could be perturbatively constructed near the symmetry-breaking point. Details of this construction will be left for future studies.

The analytical construction of soliton solutions is often a precursor of the subsequent linear stability analysis of these solitons. Thus, the results of this article could be helpful for the analytical stability investigations of solitons in Wadati-type complex potentials.

\section*{Acknowledgement}

This material is based upon work supported by the Air Force Office of Scientific Research under award number FA9550-18-1-0098 and the National Science Foundation under award number DMS-1910282.

\end{document}